\documentclass[showpacs,prd,twocolumn,aps]{revtex4}

\usepackage{amsmath}
\usepackage{amssymb}
\usepackage{latexsym}
\usepackage{amsfonts}
\usepackage{epsfig}
\usepackage{psfrag}
\usepackage{graphicx}

\newcommand{\vel}{{\mathfrak v}}

\newcommand{\inner}[2]{\left(#1|#2\right)}

\newcommand{\nn}{\nonumber\\}

\newcommand{\bea}{\begin{eqnarray}}
\newcommand{\ea}{\end{eqnarray}}
\newcommand{\eea}{\end{eqnarray}}
\newcommand{\ord}{\,{\cal O}}

\begin{document}

\title{On Quantum Correlations across the Black Hole Horizon}

\author{Ralf Sch\"utzhold$^{1,*}$ and William G.~Unruh$^{2,+}$}

\affiliation{
$^1$Fakult\"at f\"ur Physik, Universit\"at Duisburg-Essen,
D-47048 Duisburg, Germany
\\
$^2$Canadian Institute for Advanced Research Cosmology and Gravity
Program
\\
Department of Physics and Astronomy, University of British Columbia,
Vancouver B.C., V6T 1Z1 Canada
\\
$^*${\tt ralf.schuetzhold@uni-due.de},
$^+${\tt unruh@physics.ubc.ca}
}

\begin{abstract}
Inspired by the condensed-matter analogues of black holes, we study the
quantum correlations across the event horizon reflecting the entanglement
between the outgoing particles of the Hawking radiation and their
in-falling partners.
For a perfectly covariant theory, the total correlation is conserved
in time and piles up arbitrary close to the horizon in the past, where
it merges into the singularity of the vacuum two-point function at the
light cone.
After modifying the dispersion relation (i.e., breaking Lorentz invariance)
for large $k$, on the other hand, the light cone is smeared out and the
entanglement is not conserved but actually created in a given rate per
unit time.
\end{abstract}

\date{\today}


\maketitle

\section{Introduction}

One of the remaining mysteries of modern physics is the question of why
black holes seem to behave as thermal objects, described by the Hawking
temperature \cite{hawking}.
The thermal nature of black holes entails important concepts such as
black hole entropy \cite{thermo} and the information ``paradox''.
It is now widely believed that understanding the origin of thermality
would be a major step towards unifying gravity and quantum theory.

Particularly puzzling is the fact that this thermal nature is apparently
not caused by some sort of equilibration process, but by a dynamical
quantum mechanism.
In the semi-classical description, the state of the quantum fields
propagating in the gravitational background is still a pure state.
The thermal nature of the Hawking radiation is explained by the
quantum correlations (across the horizon) between the Hawking
particles and their infalling partners \cite{unruh}.

For Hawking radiation itself, the origin of the created particles
and its robustness against modifications of the microscopic structure
have been studied in many publications -- often inspired by the
condensed-matter analogues of black holes (``dumb holes''),
see, e.g., \cite{unruh-prl,universality,origin}.
In contrast, the correlations across the horizon, their origin and
dependence on the microscopic structure have been studied in far
less detail \cite{Balbinot}.
%

\section{Two-point function}

Exploiting the fact that Hawking radiation is basically a
one-dimensional effect and applies to all fields, we study
a massless scalar field $\phi$ in a 1+1 dimensional space-time
described by the Painlev{\'e}-Gullstrand-Lema{\^\i}tre
coordinates ($\hbar=c=G_{\rm N}=k_{\rm B}=1$)
\bea
ds^2
&=&
dt^2-[dx+\vel(x)dt]^2
\nn
&=&
[1-\vel^2(x)]dt^2-2\vel(x)dt\,dx-dx^2
\,,
\ea
where $\vel(x)$ can be visualized as the local velocity
of freely falling frames as measured from infinity.
The Schwarzschild metric is obtained by $\vel=\sqrt{2M/r}$,
but we shall consider arbitrary profiles $\vel(x)$.
In the standard manner, we introduce light-cone variables
$u$ and $v$ via
\bea
ds^2
&=&
[1-\vel^2(x)]
\left(dt-\frac{dx}{1-\vel(x)}\right)
\left(dt+\frac{dx}{1+\vel(x)}\right)
\nn
&=&
[1-\vel^2(x)]\,du\,dv
\,,
\ea
with $u$ diverging at the (future) event horizon $\vel=1$.
The past horizon with $\vel=-1$ corresponds to
$v\uparrow\infty$.
For future convenience, we introduce the tortoise
coordinate $dx_*=dx/[1-\vel(x)]$ with $u=t-x_*$.
After the standard transformation to regular coordinates
for $\vel<1$ (i.e., outside the horizon)
\bea
U=-\frac1\kappa\,e^{-\kappa u}
\,,\quad
V=\frac1\kappa\,e^{\kappa v}
\ea
where $\kappa=(d\vel/dx)_{\rm horizon}$ is the surface gravity,
and analytic continuation beyond horizon $\vel>1$ where $U>0$,
we obtain the line element with a regular conformal factor
\bea
ds^2=\mho^2_{\rm regular}(UV)\,dU\,dV
\,.
\ea
Due to the conformal invariance of the massless scalar field in
1+1 dimensions, we may directly read off the two-point function(s).
In the Boulware state \cite{Boulware}, which is the ground state
of the Hamiltonian generating the $t$-evolution
(i.e., of all stationary observers),
it behaves as $\ln(\Delta u\Delta v)$.
However, this quantity is clearly divergent at both horizons,
$u\uparrow\infty$ and $v\uparrow\infty$.
Black hole evaporation is described by the Unruh state, which is
regular across the black hole (future) horizon at $U=0$ leading
to the two-point function $\propto\ln(\Delta U\Delta v)$.
In this case, the ingoing $v$-modes are in their ground state.
The Israel-Hartle-Hawking state \cite{IHH} is regular across
both horizons $U=0$ and $V=0$ and is thermal in both directions.
Its  two-point function reads
(up to an undetermined constant reflecting the IR-divergence
in 1+1 dimensions)
\bea
\langle\phi(U,V)\phi(U',V')\rangle
=
-\frac{1}{4\pi}\,\ln(\Delta U\Delta V)
\,.
\ea
where $\Delta U=U-U'$ and $\Delta V=V-V'$.
Since the ingoing ($V,v$) sector is decoupled from the outgoing
($U,u$) sector (in this 1+1 dimensional set-up), we focus on the
relevant $\Delta U$ part $\langle\dots\rangle_U$
describing the Hawking radiation and the
in-falling partner particles in the following.
If $U$ and $U'$ lie on different sides of the horizon $UU'<0$,
we obtain
\bea
\langle\phi(U,V)\phi(U',V')\rangle_U
=
-\frac{1}{4\pi}\,
\ln\left(e^{-\kappa u}+e^{-\kappa u'}\right)
\,.
\ea
In this form, correlations across the horizon do not become
particularly apparent, but if we calculate, for example,
the correlator
\bea
\label{correlator-dot}
\langle\dot\phi(t,x)\dot\phi(t',x')\rangle_U
&=&
-\frac{1}{4\pi}\,\partial_t\,\partial_{t'}\,
\ln\left(e^{-\kappa u}+e^{-\kappa u'}\right)
\nn
&=&
\frac{\kappa^2}{16\pi}\,
\frac{1}{\cosh^2(\kappa[u+u']/2)}
\,,
\ea
with $u+u'=t+t'-x_*-x_*'$, we see that it has a peak 
if we regard the the Hawking particle at  ($x_*$) at time $t$  
and its in-falling partner is at  ($x_*'$)
at time $t'$ \cite{Balbinot}.

Due to $u\uparrow\infty$ at the horizon, this correlator vanishes
where $\vel=1$.
This reflects the critical slow-down (of the $u$-modes) at the
horizon in terms of the $t$-coordinate.
However, for other quantities such as the momentum density
$\Pi=(\partial_t-\vel\partial_x)\phi$, the correlator does
not vanish when approaching the horizon
\bea
\label{correlator-Pi}
\langle\Pi(t,x)\Pi(t',x')\rangle_U
&=&
-\frac{1}{4\pi}\,\partial_x\,\partial_{x'}\,
\ln\left(e^{-\kappa u}+e^{-\kappa u'}\right)
\nn
&=&
\frac{\kappa^2[1-\vel(x)]^{-1}[1-\vel(x')]^{-1}}
{16\pi\cosh^2(\kappa[u+u']/2)}
\,,
\ea
in view of $\partial_{x}u=-1/(1-\vel)$.
We observe that the $\dot\phi$ correlator across the horizon is always
positive whereas the $\Pi$ correlator is negative.
This can be explained by the fact that the particles of the Hawking
radiation and their in-falling partners have the same conserved frequency
$\omega$ measured with respect to the time $t$ but their $k$-values
have the opposite sign.
Note that $\omega$ and $k$ should not be confused with energy and
momentum: The Hawking particle has positive energy and momentum
whereas the in-falling partner has negative energy but positive
momentum, see, e.g., \cite{Schutzhold+Maia}.

\section{Integrals}

Since the $\Pi$-correlator is total derivative, we may easily
derive the total correlation integrated from the horizon at
$x=0$ up to spatial infinity
\bea
\label{conserved}
\int\limits_0^\infty dx\,\langle\Pi(t,x)\Pi(t',x'<0)\rangle_U
=
\frac{\kappa}{4\pi}\,\frac{1}{1-\vel(x')}
\,,
\ea
and we find that it is independent of $t-t'$.
Consequently, for each point $x'<0$ inside the horizon (at $x=0$),
the integral of the correlations across the horizon is conserved,
i.e., independent of the time-slice.
In the far future $t\uparrow\infty$, the Hawking particles carry the
correlations to spatial infinity -- as one would expect.
In the past $t\downarrow-\infty$, however, the correlations pile
up near the horizon where $u\approx t-\kappa^{-1}\ln(\kappa x)$
\bea
\langle\Pi(t\downarrow-\infty,x\downarrow0)\Pi(t',x'<0)\rangle_U
\approx
\nn
\frac{\kappa}{16\pi}\,
\frac{[1-\vel(x')]^{-1}}{\cosh^2(\kappa[u+u']/2)}
\,\frac{1}{x}
\,,
\ea
i.e., they are concentrated in a small spatial volume and
have a large amplitude (in order to keep the integral constant).
As we shall see in the next Section, these piled-up correlations
merge into the singularity of the two-point function at the light
cone (which approaches the horizon) and become virtually
indistinguishable from the quantum vacuum fluctuations.
For a modified dispersion relation, this picture changes
drastically, see Section~\ref{Dispersion}.

Similarly, we may evaluate the time-integral
\bea
\label{t-int-Pi}
\int\limits_{-\infty}^{+\infty} dt\,
\langle\Pi(t,x)\Pi(t',x')\rangle_U
=
\frac{\kappa}{4\pi}\,
\frac{1}{1-\vel(x)}\,\frac{1}{1-\vel(x')}
\,.
\ea
In contrast to the $x$-integral above, this result will survive
for a modified dispersion relation (apart from some corrections
at short length scales, see Section~\ref{Dispersion}).
Of course, we may also derive the time-integral for
\bea
\label{t-int-dot}
\int\limits_{-\infty}^{+\infty} dt\,
\langle\dot\phi(t,x)\dot\phi(t',x')\rangle_U
=
\frac{\kappa}{4\pi}
\,.
\ea
The difference between the two expressions (\ref{t-int-Pi})
and (\ref{t-int-dot}) can again be traced back to the fact that
the Hawking particles and their in-falling partners have the
same $\omega$, but different $k$ depending on their positions
$x$ and $x'$ (gravitational red-shift $\omega=[1-\vel]k$).

\section{Analytically solvable example}

It might be illustrative to apply the above formulae
(which are valid for arbitrary profiles $\vel$) to some simple
example which allows us to write down closed expressions.
To this end, let us choose the velocity profile
\bea
\vel(x)=1-\frac{\kappa}{\gamma}\tanh(\gamma x)
\,,
\ea
where $\gamma$ is some parameter and $\kappa$ the surface gravity.
In this case, the light cone coordinates read
\bea
u=t-\frac1\kappa\,\ln[\sinh(\gamma x)]
\,\leadsto\,
U=-\frac1\kappa\,e^{-\kappa t}\sinh(\gamma x)
\,,
\ea
and the correlator for all $x$ and $x'$ becomes
\bea
\langle\dot\phi(t,x)\dot\phi(t,x')\rangle_U
=
-\frac{\kappa^2}{4\pi}\,
\frac{\sinh(\gamma x)\sinh(\gamma x')}
{[\sinh(\gamma x)-\sinh(\gamma x')]^2}
\,.
\ea
This function is positive on opposite sides of the horizon $xx'<0$,
negative when $x$ and $x'$ lie on the same side $xx'>0$, and goes to
zero if one of the two points approaches the horizon at $x=0$
(critical slow-down).
It reproduces the usual $1/(x-x')^2$ singularity at $x=x'$ and
vanishes asymptotically $x\to\pm\infty$ and $x'\to\pm\infty$.
The correlations across the horizon manifest themselves in the
global maximum at $x=-x'$ with $t=t'$.

\begin{figure}[ht]
\includegraphics[width=0.48\textwidth]{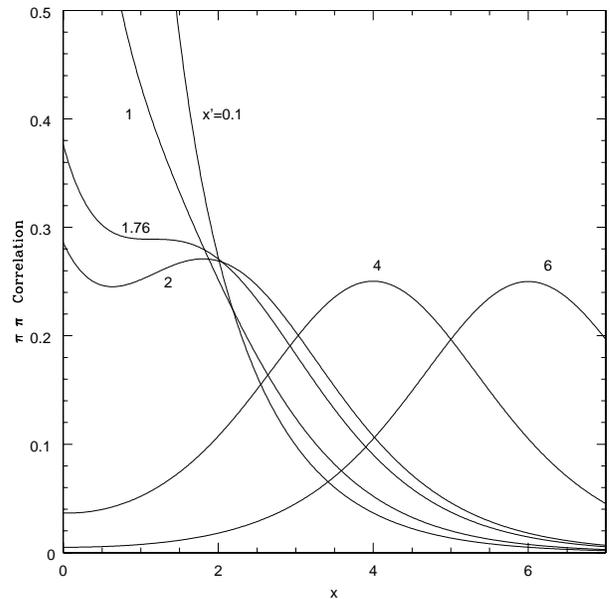}
\caption{\label{figure}Plot of the correlator (\ref{Pi-Pi-gamma})
as a function of position $x$ outside the horizon for various values
of the partner position $x'$ inside the horizon
(both in units of $\gamma$).
For large values of $x'$ (i.e., far from the horizon) the correlation
as a function of $x$ is simply a bump of width about 2 and constant
height which is located at $x=x'$.
For small $x'$ it merges into the general vacuum fluctuations near
the horizon, i.e., the $1/(x-x')^2$ singularity.
I.e., the correlation of the "Hawking" particles with their infalling
partners becomes visible only if the two points are well separated from
the horizon, even though the integral (\ref{conserved}) of the
correlation over $x$ is a constant.}
\end{figure}

Although one gets the same $1/(x-x')^2$ singularity at $x=x'$
and the same asymptotic behavior for $x\to\pm\infty$ and
$x'\to\pm\infty$, the situation is a bit different for the
canonical momentum density $\Pi=\dot\phi-\vel\phi'$
\bea
\label{Pi-Pi-gamma}
\langle\Pi(t,x)\Pi(t,x')\rangle_U
=
-\frac{\gamma^2}{4\pi}\,
\frac{\cosh(\gamma x)\cosh(\gamma x')}
{[\sinh(\gamma x)-\sinh(\gamma x')]^2}
\,.
\ea
First, this correlator is negative everywhere and does not vanish
at the horizon.
Second, we get a local minimum at $x=-x'$ only if
$|x|>x_*\approx1.8/\gamma$, i.e., far enough away from the horizon.
Close to the horizon, the local minimum merges into the light-cone
singularity and disappears, see Figure~\ref{figure}.
This absence of structure near the horizon can be interpreted as
further confirmation of the picture that the Hawking radiation is
not created very close to the horizon but rather in a region of
finite spatial extent $\ord(1/\gamma)$.
Note that this length scale $1/\gamma$ could in principle be quite
different from the scale set by the Hawking temperature, i.e.,
the surface gravity $\kappa$.

It should also be mentioned that the $\dot\phi$-correlator {\em does}
have a maximum arbitrarily close to the horizon, but the maximum value
becomes very small.
Consistent with this observation, the correlation conservation law
(\ref{conserved}) applies to the $\Pi$-correlator -- for the
$\dot\phi$-correlator, one would need an additional integrating
factor $1/[1-\vel(x)]$.

\section{Modified Dispersion}\label{Dispersion}

In the previous Sections, we found the infinite pile up of correlation
close to the horizon in the past, such that the correlations constantly
emerge out of the singularity at the light-cone (which approaches the
null surface of the horizon in the past).
However, modifying the dispersion relation at large $k$
(e.g., motivated by the condensed-matter analogues of black holes
\cite{unruh-prl,universality,origin}),
we expect this behavior to change:
First, a modified dispersion relation smears out the light cone
such that the two-point function typically becomes regular everywhere
except at the space-time coincidence point $x=x'$ and $t=t'$.
Second, tracing the particles of the Hawking radiation back in time,
they do not originate from an small vicinity of the horizon in the
presence of a modified dispersion relation
\cite{universality,origin}.

In order to deal with a solvable example, let us switch to the
Eddington-Finkelstein coordinates $(v,r)$
\bea
ds^2=\left(1-\frac{2M}{r}\right)dv^2-2dv\,dr
\,,
\ea
and modify the dispersion relation by inserting a general function
$f(\partial_r^2)$ containing higher-order spatial derivatives into
the corresponding action for a scalar field $\phi$
\bea
{\cal L}=
-(\partial_v\phi)\partial_r\phi-
\frac{(\partial_r\phi)}{2}
\left[1-\frac{2M}{r}+f(\partial_r^2)\right]
\partial_r\phi
\,.
\ea
This action allows us to define a conserved inner product
\bea
\label{inner}
\inner{\phi_1}{\phi_2}
=
i\int d\Sigma^\mu\,
\phi_1^*\stackrel{\leftrightarrow}{\partial}_\mu\phi_2
=
i\int dr\,\phi_1^*\stackrel{\leftrightarrow}{\partial}_r\phi_2
\,,
\ea
where $\phi_1^*\stackrel{\leftrightarrow}{\partial}_r\phi_2=
\phi_1^*\partial_r\phi_2-\phi_2\partial_r\phi_1^*$.
The momentum density $\Pi=-\partial_r\phi$ satisfies the wave
equation
\bea
\left(2\partial_v+
\partial_r\left[1-\frac{2M}{r}+f(\partial_r^2)\right]
\right)\Pi=0
\,,
\ea
which is of first order in $\partial_v$ and hence
automatically selects the outgoing $u$-sector only.
In a stationary state, the two-point function can be
Fourier expanded via
\bea
\label{Fourier}
\langle\Pi(v,r)\Pi(v',r')\rangle=
\int d\omega\,e^{-i\omega(v-v')}\,g_\omega(r,r')
\,,
\ea
where $g_\omega(r,r')$ solves the ordinary differential equation
\bea
\left(-2i\omega+
\partial_r\left[1-\frac{2M}{r}+f(\partial_r^2)\right]
\right)g_\omega(r,r')=0
\ea
for $r$ and the same for $r'$ with $+2i\omega$.
Assuming that this differential equation together with the
asymptotic conditions (freely falling ground state for large
$k$ at all positions $r$) uniquely determines the $r$-dependence
of $g_\omega(r,r')$ (and thus the same for $r'$), we find that
$g_\omega(r,r')$ factorizes
\bea
\label{factorize}
g_\omega(r,r')=h_\omega(r)\,h^*_\omega(r')
\,.
\ea
Here, we are interested in the region near the horizon
(where the pile-up of correlation occurred) and thus
we employ the near-horizon approximation
\bea
1-\frac{2M}{r}=2\kappa x+\ord(\kappa^2x^2)
\,,
\ea
resulting in the (approximate) differential equation 
which can be solved via a Fourier-Laplace transformation
\bea
\left(-2i\omega+
\partial_x
\left[2\kappa x+f(\partial_x^2)\right]
\right)h_\omega(x) &=& 0
\nn
\left(-2i\omega+
ik
\left[2\kappa i\partial_k+f(-k^2)\right]
\right)\tilde h_\omega(k) &=& 0
\,.
\ea
%
%
%
This is now a first-order ordinary differential equation
in $k$ and its general solution can be written as
\bea
h_\omega(x)=\int\limits_{\mathfrak C} dk\,k^{-i\omega/\kappa}
\exp\left\{ikx-\frac{i}{\kappa}F(k^2)\right\}
\,,
\ea
where ${\mathfrak C}$ is an appropriate contour in the
complex plane and $dF/dk=f(-k^2)/2$ accounts for the modified
dispersion relation.
In order to determine the correct integration contour in the
complex plane, we have to study different choices for the branch
cut from $k^{-i\omega/\kappa}$.
If the branch cut lies in the upper complex half plane $\Im(k)>0$,
the solution connects the final Hawking mode to the initial
positive/negative (pseudo) norm modes \cite{origin}
\bea
\label{hawking}
\alpha_\omega
\phi_\omega^{k^+}(x)
+
\beta_\omega
\phi_\omega^{k^-}(x)
\to
\phi_\omega^{\rm Hawking}(x>0)
\,,
\ea
where $\phi_\omega^{k^+}(x)$ has positive (pseudo) norm
(\ref{inner}) and $\phi_\omega^{k^-}(x)$ negative (pseudo) norm.
On the other hand, the branch cut in the lower complex
half plane $\Im(k)<0$ connects the final mode of the in-falling
partners to the initial positive/negative (pseudo) norm modes
\bea
\label{partner}
\tilde\alpha_\omega
\phi_\omega^{k^-}(x)
+
\tilde\beta_\omega
\phi_\omega^{k^+}(x)
\to
\phi_\omega^{\rm partner}(x<0)
\,.
\ea
Assuming that the quantum state we have corresponds to the freely
falling ground state for large $k$, the asymptotic condition for
$h_\omega(x)$ implies that it has no contribution from
$\phi_\omega^{k^-}(x)$.
Thus, we take a suitable linear combination of the two solutions
(\ref{hawking}) and (\ref{partner}) with the branch cut in the
upper and lower complex half-plane, respectively, which yields
\bea
\label{linear-combination}
h_\omega(x)
=
{\cal N}_\omega\int\limits_0^\infty dk\,
k^{-i\omega/\kappa}
\exp\left\{ikx-\frac{i}{\kappa}F(k)\right\}
\,,
\ea
where ${\cal N}_\omega$ is a normalization factor.
This expression is quite natural since the boundary condition
(freely falling ground state for large $k$) implies that
$h_\omega(x)$ only contains positive $k$-values with a
positive (pseudo) norm (\ref{inner}).
Coarse-graining over large length scales, we do not see the
impact of $F(k)$ and this function behaves as
$h(x)\sim|x|^{i\omega/\kappa-1}$, but on short distances,
it also contains the rapidly oscillating in-mode
$\phi_\omega^{k^+}(x)$.

\begin{widetext}

Now we are in the position to study the full correlator.
Inserting (\ref{linear-combination}) into
(\ref{Fourier}) and (\ref{factorize}), the total expression reads
\bea
\langle\Pi(v,x)\Pi(v',x')\rangle=
\int\limits_{-\infty}^{+\infty} d\omega\,e^{-i\omega(v-v')}\,
\left|{\cal N}_\omega^2\right|
\int\limits_0^\infty dk\,\int\limits_0^\infty dk'\,
\exp\left\{
i\frac{\omega}{\kappa}\ln\frac{k'}{k}
+ikx-ik'x'
-\frac{i}{\kappa}F(k)
+\frac{i}{\kappa}F(k')
\right\}
\,.
\ea
The $\omega$-integral yields the Fourier transform of
$|{\cal N}_\omega^2|$, which we denote by $\widetilde{\cal N}$.
Finally, introducing the new variable $\chi=e^{-\kappa(v-v')}k/k'$,
the integrated correlation across the horizon in analogy to
(\ref{conserved}) yields
\bea
\label{conserved-disp}
\int\limits_0^\infty dx\,
\langle\Pi(v,x)\Pi(v',x'<0)\rangle
=
\int\limits_0^\infty dk'\,
\exp\left\{
-ik'x'+\frac{i}{\kappa}F(k')
\right\}
\times
\int\limits_0^\infty\frac{d\chi}{\chi}\,
\widetilde{\cal N}(\ln\chi)
\exp\left\{
-\frac{i}{\kappa}F\left(\chi k'e^{-\kappa(v-v')}\right)
\right\}
\,.
\ea
%

\end{widetext}

For a given point $x'<0$ inside the horizon, Eq.~(\ref{conserved-disp})
yields the total correlation between that point $x'<0$ and all positions
$x$ outside the horizon up to spatial infinity.
Setting $F=0$, we rederive the result (\ref{conserved})
for $\vel(x)=\kappa x$.
In the far future $(v-v')\uparrow\infty$, the exponential pre-factor
$e^{-\kappa(v-v')}$ in Eq.~(\ref{conserved-disp}) vanishes and thus
the $\chi$-integral becomes independent of $k'$.
The remaining $k'$-integral then just yields $h^*_{\omega=0}(x')$.
After coarse-graining over large length scales, this scales as
$1/x'$.
As a result, we get basically the same conservation law as in
(\ref{conserved}).
This is quite natural since it just reflects that fact that the
Hawking particles carry the correlation away to spatial infinity.

In the far past $(v-v')\downarrow-\infty$, on the other hand,
the exponential pre-factor $e^{-\kappa(v-v')}$ diverges and thus
the $k'$-integral in Eq.~(\ref{conserved-disp}) is exponentially
suppressed due to the rapidly oscillating phase
$F(\chi k'e^{-\kappa(v-v')})$.
This can most easily be seen by changing the integration
variable to $k''=k'e^{-\kappa(v-v')}$.
In the resulting double integral over $k''$ and $\chi$,
the first exponent in Eq.~(\ref{conserved-disp}) can be
neglected and the integral scales as $e^{\kappa(v-v')}$.
Consequently, in contrast to the perfectly covariant case
(which implies an unbounded red-shift near the horizon),
the total correlation is not conserved in this case
(i.e., it does not pile up at the horizon)
but created at finite times.

Note that the integral over the full $x$-interval
(i.e, from $-\infty$ to $+\infty$) vanishes,
since $\Pi$ is a spatial derivative.
Even with a modified dispersion relation,
the time-integral factorizes exactly
\bea
\int\limits_{-\infty}^{+\infty} dv\,
\langle\Pi(v,x)\Pi(v',x')\rangle
=
2\pi h_{\omega=0}(x)h^*_{\omega=0}(x')
\,,
\ea
and, after coarse-graining over large length scales, it behaves
as $1/|xx'|$.

\section{Conclusions}

We have studied the evolution of the quantum correlations across the
black hole horizon.
Since the quantum state under consideration is a pure state, all
correlations imply entanglement and thus entanglement entropy etc.
Both, the $\dot\phi$ and the $\Pi$ correlator across the horizon
(\ref{correlator-dot}) and (\ref{correlator-Pi}), possess a peak
at $u=-u'$ if we are far enough away from the horizon.
For black hole analogues (``dumb holes'') in Bose-Einstein condensates,
the scalar field $\phi$ reflects the phase fluctuations while the
momentum density $\Pi$ corresponds to the density fluctuations
$\delta\varrho$, cf.~\cite{Balbinot}.

For the $\Pi$ correlator, we found a conservation law
(\ref{conserved}) for the total correlation across horizon in a
perfectly covariant theory (up to arbitrarily small length scales).
This means that the correlation to be carried away by the
Hawking particles in the future must pile up arbitrarily close
to the horizon in the past.
However, below a minimum length scale set by the geometry
(not necessarily the surface gravity), this piled-up
correlation becomes virtually indistinguishable from the vacuum
singularity of the two-point function at the light cone,
see Figure~\ref{figure}.

After modifying the microscopic structure via introducing a
non-linear dispersion relation at short distances, this picture
changes drastically:
In this case, the correlation carried away by Hawking radiation
in the future cannot be traced back to arbitrarily early times
and a small vicinity of the horizon.
As a result, the entanglement is not conserved but actually
created dynamically at a finite time.
These findings further support the view that Hawking radiation
is not created at arbitrarily small length scales but at
finite distances and could be relevant for the black hole
information ``paradox'' etc.

\bigskip

\section*{Acknowledgement}

This work was supported by the Emmy-Noether Programme of the German
Research Foundation (DFG) under grant \# SCHU~1557/1-2,3; as well as
the Canadian Institute for Advanced Research; and the
Natural Science and Engineering Research Council of Canada.


\end{document}